\documentclass[letterpaper]{article} 

\ifdefined\aaaianonymous
    \usepackage[submission]{aaai2026}  
\else
    \usepackage{aaai2026}              
\fi

\usepackage{times}  
\usepackage{helvet}  
\usepackage{courier}  
\usepackage[hyphens]{url}  
\usepackage{graphicx} 
\usepackage{natbib}  
\usepackage{caption} 
\usepackage{multirow}
\usepackage{arydshln}
\usepackage{subcaption}
\usepackage{amsmath}
\usepackage{booktabs}
\usepackage{amssymb}

\usepackage{algorithm}
\usepackage{algorithmic}

\usepackage{newfloat}
\usepackage{listings}
\DeclareCaptionStyle{ruled}{labelfont=normalfont,labelsep=colon,strut=off} 
\floatstyle{ruled}
\newfloat{listing}{tb}{lst}{}
\floatname{listing}{Listing}

\ifdefined\aaaianonymous
    \title{AV-SSAN: Audio-Visual Selective DoA Estimation through Explicit Multi-Band Semantic-Spatial Alignment}
\else
    \title{AV-SSAN: Audio-Visual Selective DoA Estimation through Explicit Multi-Band Semantic-Spatial Alignment}
\fi

\author{
    Yu Chen\textsuperscript{\rm 1,5},
    Hongxu Zhu\textsuperscript{\rm 2},
    Jiadong Wang\textsuperscript{\rm 3},
    Kainan Chen\textsuperscript{\rm 4},
    Xinyuan Qian\textsuperscript{\rm 5}
}
\affiliations{
    \textsuperscript{\rm 1}School of Data Science, The Chinese University of Hong Kong (Shenzhen), China\\
    \textsuperscript{\rm 2}Fano, Hong Kong\\
    \textsuperscript{\rm 3}Technical University of Munich, Germany\\
    \textsuperscript{\rm 4}Eigenspace GmbH, Germany\\
    \textsuperscript{\rm 5}University of Science and Technology Beijing (USTB), China\\
    
}

\usepackage{bibentry}

\begin{document}

\maketitle

\begin{abstract}
Audio-visual sound source localization (AV-SSL) estimates the position of sound sources by fusing auditory and visual cues. Current AV-SSL methodologies typically require spatially-paired audio-visual data and cannot selectively localize specific target sources. To address these limitations, we introduce Cross-Instance Audio-Visual Localization (CI-AVL), a novel task that localizes target sound sources using visual prompts from different instances of the same semantic class. CI-AVL enables selective localization without spatially paired data. To solve this task, we propose AV-SSAN, a semantic-spatial alignment framework centered on a Multi-Band Semantic-Spatial Alignment Network (MB-SSA Net). MB-SSA Net decomposes the audio spectrogram into multiple frequency bands, aligns each band with semantic visual prompts, and refines spatial cues to estimate the direction-of-arrival (DoA). 
To facilitate this research, we construct VGGSound-SSL, a large-scale dataset comprising 13,981 spatial audio clips across 296 categories, each paired with visual prompts. AV-SSAN achieves a mean absolute error of 16.59° and an accuracy of 71.29\%, significantly outperforming existing AV-SSL methods. Code and data will be public upon acceptance.

\end{abstract}

\ifdefined\aaaianonymous
\else
\fi







\section{Introduction}
Sound Source Localization (SSL) estimates the DoA of sound sources from multichannel audio. Classical methods, such as GCC-PHAT~\cite{GCC}, MUSIC~\cite{MUSIC}, and SRP-PHAT~\cite{SRPPHAT}, rely on spatial spectrum estimation or time-delay analysis. While effective in controlled settings, they degrade severely in the presence of multiple sources, high background noise, or strong reverberation.


\begin{figure}
    \centering
    \includegraphics[width=\linewidth]{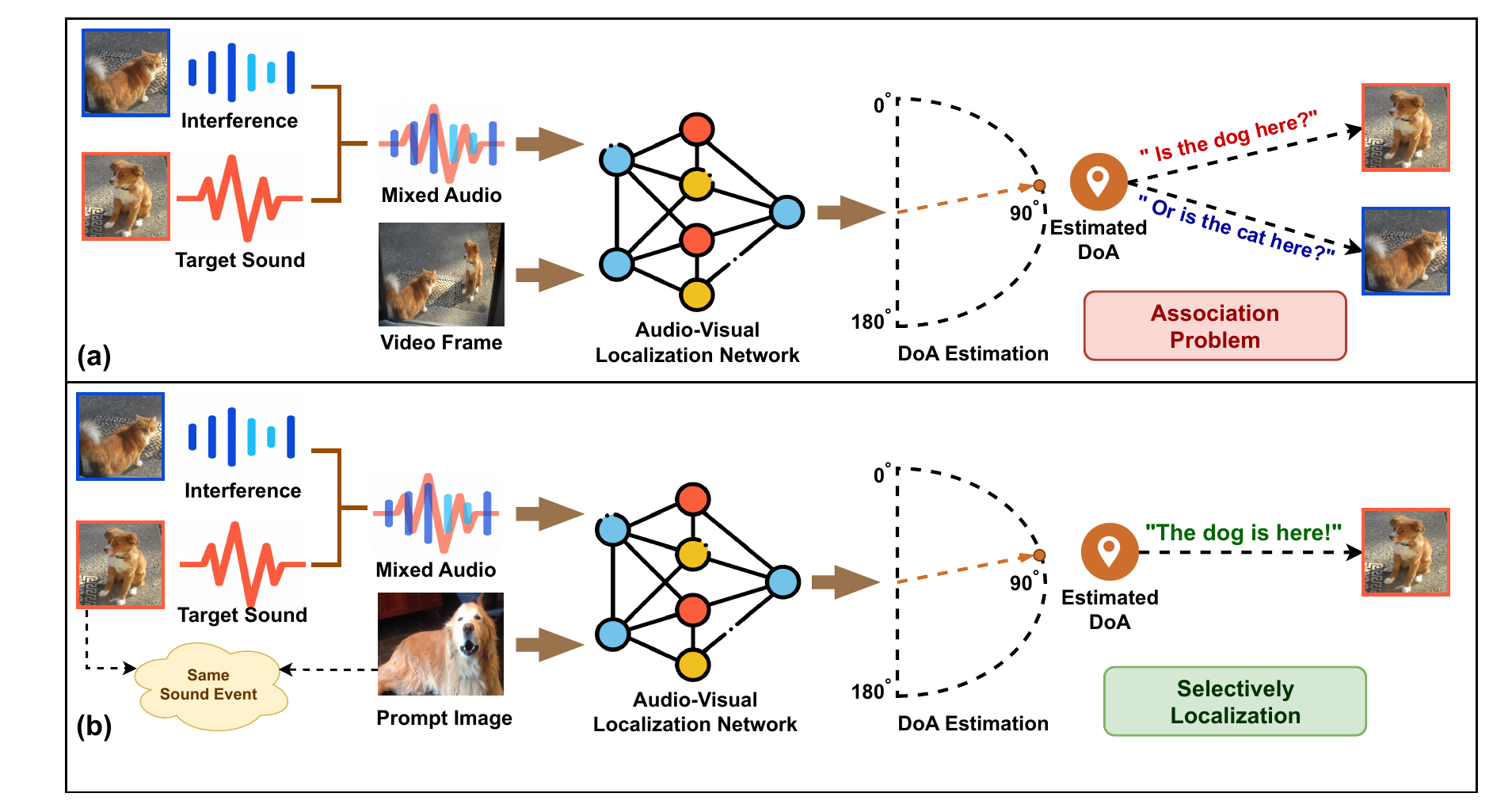}
    \caption{Comparison between conventional AV-SSL and our proposed CI-AVL task. (a) Traditional AV-SSL requires spatially-paired data but cannot selectively localize targets. (b) Our CI-AVL enables selective localization using semantic visual prompts from different instances, eliminating the need for spatial pairing.}
    \label{intro}
\end{figure}

To overcome these limitations, recent works leverage deep neural networks (DNNs) to learn spatial representations either to enhance classical methods~\cite{gccmlp,music1,srp2} or in an end-to-end form~\cite{cnndoa,cnnlstm_doa,tfmamba}. However, these audio-only models are fragile in acoustically challenging scenes or when the target sound source is inactive.

Visual modalities offer a complementary cue to SSL. Recent AV-SSL frameworks~\cite{qianmulti,avmlp,dgb, CMAF,lipnet,avseld,AVST,TAVF} integrate spatial audio and visual context to improve localization performance.
However, these methods have two major limitations: (1) They require tightly spatially-aligned audio-visual inputs where the visible object directly corresponds to the sounding source, a condition rarely met in real-world data. (2) They localize all active sources but lack the ability to selectively localize a specific target source of interest.

To address these limitations, we propose a new task: \textbf{Cross-Instance Audio-Visual Localization (CI-AVL)}. 
CI-AVL aims to localize a target source using a visual prompt derived from a different instance of the same semantic class (e.g., localizing a barking dog with an image of another dog). As depicted in Figure~\ref{intro}, this setting enables selective localization without requiring explicitly paired audio-visual data. However, CI-AVL presents a unique challenge: the visual prompt is only semantically associated with the sound source but spatially unrelated, making direct fusion strategies employed by prior AV-SSL methods ineffective.

This challenge stems from a fundamental semantic-spatial misalignment: existing methods primarily focus on spatial alignment (``where"), while overlooking semantic alignment (``what") between audio and visual modalities. In contrast, human perception follows a hierarchical process: we first semantically recognize the sound source (``what"), then localize its position (``where") \cite{semantic1,semantic2}. Inspired by this, we hypothesize that effective target-aware localization requires a two-stage alignment mechanism: (1) aligning the visual prompt and audio semantically to isolate the target's identity, and (2) localizing the target source conditioned on the identity.

To this end, we introduce AV‑SSAN, which first semantically aligns a cross-instance visual prompt with mixed audio and then spatially localizes the target source. Inspired by the frequency-dependent characteristics of spatial hearing, AV‑SSAN incorporates an MB‑SSA Net. It decomposes the spectrogram into different frequency resolutions, semantically aligns each band with the visual prompt, and then fuses them via an attention-guided refiner to yield DoA estimates.

In addition, we introduce VGGSound-SSL, a large-scale dataset constructed from VGGSound~\cite{vggsound}. It contains 13,981 spatial audio clips across 296 categories, each paired with semantically matched visual prompts. 
Extensive experiments show that AV-SSAN outperforms other AV-SSL baselines. Our contributions are listed as follows:
\begin{itemize}
    \item We formulate CI-AVL, a novel task designed to enable selective localization of a target source. It utilizes semantic visual prompts derived from a different instance of the same class, thereby relaxing the requirement of explicitly paired audio-visual data.
    
    \item We propose AV-SSAN, an innovative framework that performs explicit Semantic-Spatial Alignment, enabling identity-aware localization by bridging visual semantics with spatial audio features.
    
    \item We propose an MB-SSA Net module, which introduces frequency-aware modeling into the alignment process. It mimics the frequency-dependent nature of spatial hearing, using a tri-band decomposition design with semantic-guided band fusion and spatial refinement.
    
    \item We construct VGGSound-SSL, a large-scale AV-SSL dataset comprising 13,981 spatial audio clips across 296 sound event categories paired with semantic visual prompts. This dataset offers a valuable benchmark for future research in identity-aware localization.
\end{itemize}

\section{Related Work}
\subsection{Audio-only SSL}
\begin{figure*}[!htb]
    \centering
    \includegraphics[width=0.82\linewidth]{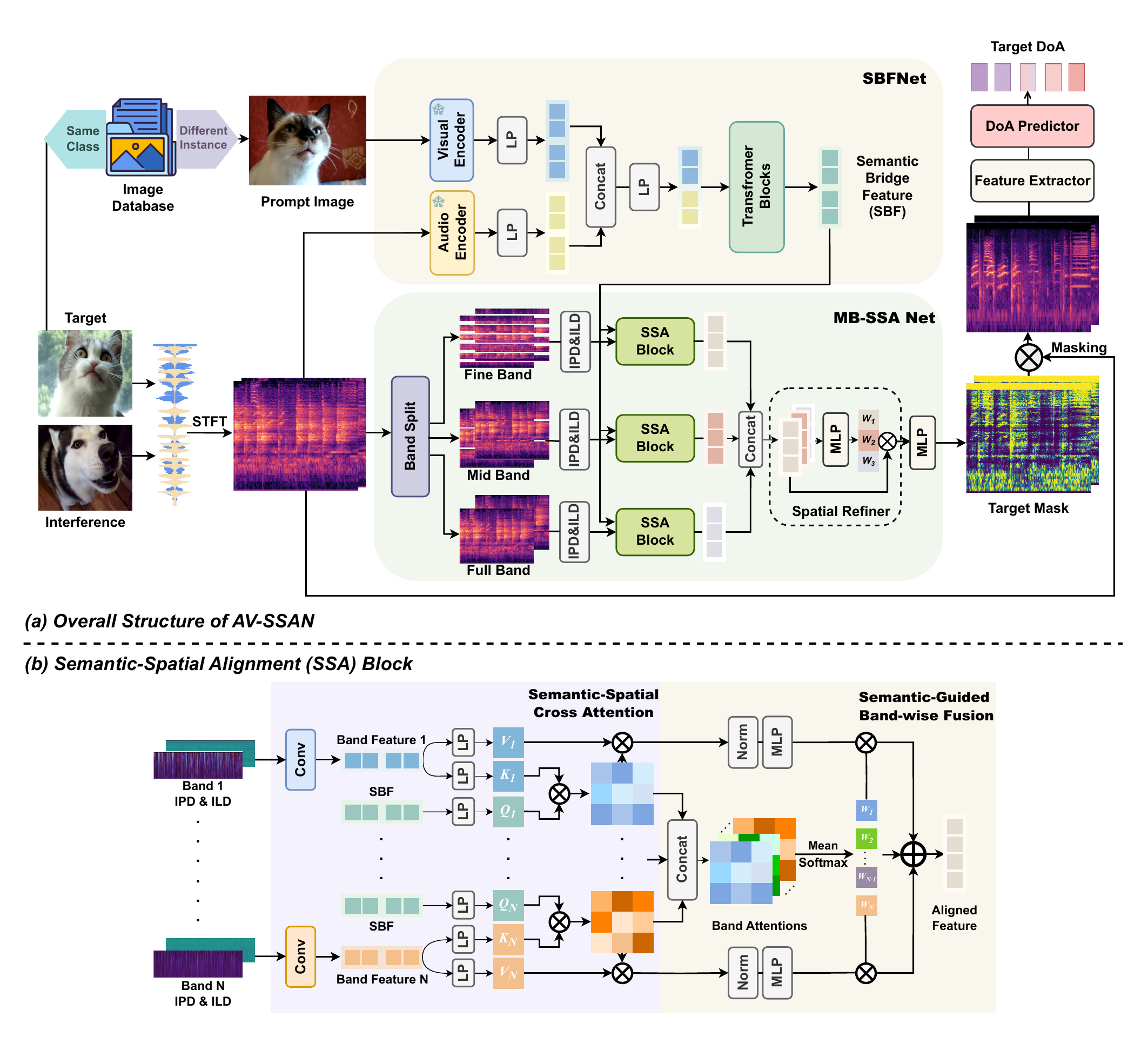}
    \caption{The architecture of (a) our proposed AV-SSAN and  (b) Semantic-Spatial Alignment (SSA) block.}
    \label{block}
\end{figure*}

Traditional SSL methods combine handcrafted spatial features with deep learning models. GCC-MLP~\cite{gccmlp} feeds GCC-PHAT features into multilayer perceptrons, while DR-MUSIC~\cite{music1} enhances covariance matrix estimation for the MUSIC algorithm. Cross3D~\cite{srp2} processes SRP-PHAT feature using 3D CNNs for localization in reverberant environments.

Fully end-to-end models directly operate on spectral inputs. CNN-based methods~\cite{cnndoa} exploit Short-Time Fourier Transform (STFT) phase cues, while SELDNet~\cite{seldnet} integrates magnitude and phase using CRNNs. Later extensions incorporate temporal modeling via LSTMs~\cite{cnnlstm_doa}, and more recently, TF-Mamba~\cite{tfmamba} replaces RNNs with Mamba~\cite{bimamba}, improving performance in complex acoustic scenes.

Despite these advances, audio-only methods lack high-level semantic understanding and fail to disambiguate overlapping sources or occluded sources.

\subsection{Audio-Visual SSL}
Recent audio-visual methods integrate visual context to enhance robustness and resolve spatial ambiguities. A prevalent strategy is fusing visual embeddings with GCC-PHAT features. Specific approaches include: MLP-AVC~\cite{qianmulti}, which models visual priors as multivariate Gaussians; AVMLP~\cite{avmlp}, which utilizes de-emphasis maps to suppress distractors; DGB~\cite{dgb}, which explores cross-modal latent spaces via generative models; CMAF~\cite{CMAF}, which employs dynamic attention for audio-visual alignment; and AVST~\cite{AVST}, which disentangles modality encoding using Vision Transformers~\cite{vit} prior to multimodal fusion.

Beyond spatial features, richer spectral representations are also employed. Some works~\cite{MultiDoA, TAVF} integrate STFTs and Gaussian-encoded visual cues through CNN or Transformers~\cite{transformer}. AV-SELD~\cite{avseld} combines Log-mel spectrograms and Intensity Vectors with visual features, fusing them via Conformer~\cite{conformer} blocks. 

Although these methods achieve promising results, their reliance on tightly synchronized audio-visual pairs limits their applicability. Moreover, they estimate all active sources indiscriminately and lack the ability to isolate a target source. These limitations stem from their emphasis on spatial correlation, while ignoring semantic alignment, which is crucial for selective localization.

\subsection{Selective and Prompt-based SSL}
Recent works explored selective localization using external semantic cues. Class-conditioned SELD models~\cite{classlocate, zerolocate} incorporate class labels to guide attention toward target categories. Text-queried SSL~\cite{textlocate} fuses textual prompts with spatial audio for selective reasoning. LocSelect~\cite{locselect} uses a reference audio to localize the corresponding speaker in mixtures. However, these approaches depend on either class labels or same-instance queries, limiting generalization to new categories or modalities.

In summary,
Audio-only methods leverage handcrafted spatial features but degrade under noisy and reverberant conditions. AV-SSL methods incorporate visual signals to enhance robustness, but they are constrained by paired data and the absence of semantic-level alignment mechanisms.
Recent advances explore selective localization using text or reference audio, but are constrained to predefined sound categories and overlook visual prompt-based conditioning. These challenges highlight the need for a semantically aware SSL framework capable of selectively extracting the target sound source, which leads to the introduction of CI-AVL.

\section{Methodology}
We introduce the AV-SSAN to address the core challenge of cross-instance semantic-spatial alignment for CI-AVL. As shown in Figure~\ref{block}, AV-SSAN comprises three modules: 1) a SBF Net that fuses visual prompts and semantic audio embeddings, 2) an MB-SSA Net that achieves multi-band semantic-spatial alignment, and 3) a DoA Prediction Module that estimates target DoA.




\begin{table*}[!htb]
\centering
\begin{tabular}{ccccccc}
    \toprule
    Dataset & Samples & Duration & Sound Event & Annotations\\
    \midrule
    AV16.3 & 43 & $\sim$2hrs & $2^{\diamondsuit}$ & DoA \\
    SSLR & 6,622 & $\sim$25hrs & $2^{\diamondsuit}$ & DoA, VAD \\
    CAV3D & 20 & $\sim$25hrs & $2^{\diamondsuit}$ & DoA, VAD \\
    AVRI & 43 & $\sim$8hrs & $2^{\diamondsuit}$ & DoA, VAD \\
    STARSS23 & 168 & $\sim$7hrs & 13 & DoA, VAD, Object Category\\
    \textbf{VGGSound-SSL (ours)} & 13,981 & $\sim$39hrs & 296 & DoA, Object Category \\
    \bottomrule
\end{tabular}
\caption{Comparison of VGGSound-SSL with existing datasets for AV-SSL. We use $\diamondsuit$ for datasets limited to male and female speech (counted as two distinct sound events). Annotations cover Object Category (sound event label), VAD (voice activity detection label), and DoA}
\label{table1}
\end{table*}

\subsection{Problem Formulation}

Given a two-channel audio signal captured by a microphone pair ${\bf x}$ and a user-specified prompt image \( \bf I \), CI-AVL aims to predict the DoA of the sound source semantically matching the prompt. We formulate this as a regression task, discretizing the DoA into 180 classes: \( \hat{\theta} = \{ j \mid 1 \leq j \leq 180, j \in \mathbb{Z} \} \). To capture spatial continuity, we model the posterior probability distribution of \( \hat{\theta} \) using a Gaussian-like vector. 
$
p(\theta) = \exp\left(-\frac{\left | \hat{\theta} - \theta \right | }{\sigma_{\theta}}\right)$ instead of one-hot encoding.
$p(\theta)$ is centered on the ground truth $\theta$ with a standard deviation $\sigma_{\theta}$. The normalization factor $\frac{1}{\sqrt{2\pi\sigma_{\theta}}}$ is omitted as it does not influence the model's predictions.

We use a deep neural network \( \mathcal{F} \) to map the multimodal inputs to the DoA distribution 
$
        \hat{p}(\theta) = \mathcal{F}({\bf I}, {\bf x}| \boldsymbol{\Omega})
        $
where \( \boldsymbol \Omega \) are learnable parameters and the DoA is determined as the direction with the highest probability
$
    \hat{\theta} = \arg\max_{\forall \theta} \hat{p}(\theta)
    $.

\subsection{SBF Net}



We first establish semantic correspondence between visual and audio modalities. We leverage a pretrained CLIP encoder~\cite{CLIP} to extract the visual prompt representation $\mathbf{F}_V \in \mathbb{R}^{d_V}$ and a VGGish network~\cite{vggish} to extract the semantic audio embedding $\mathbf{F}_A \in \mathbb{R}^{t \times d_A}$.

After temporal broadcasting and linear projection to a shared space, $\mathbf{F}^{'}_V$ and $\mathbf{F}^{'}_A$ are concatenated and processed by a Transformer encoder:
\begin{equation}
\mathbf{SBF} = \text{Transformer}(\text{LP}(\mathbf{F}'_V \textcircled{c} \mathbf{F}'_A))
\end{equation}
where $\textcircled{c}$ denotes concatenation along the feature dimension and LP is a linear projection layer.

This SBF captures the semantic information of the target sound source, serving as a semantic cue for the subsequent target source's spatial characteristics disentanglement.


\subsection{MB-SSA Net}


Auditory spatial perception is inherently frequency-dependent~\cite{spatialhear2}. Motivated by psychoacoustic principles and prior empirical findings~\cite{spatialnet}, we design MB-SSA Net to model semantic-spatial alignment across different frequency bands. The architecture comprises three parts: 1) Tri-band Spatial Feature Extraction Module, 2) Semantic-Spatial Alignment (SSA) block, and 3) Spatial Refiner.

\noindent \textbf{Tri-band Spatial Feature Extraction.} Human auditory perception distinguishes spatial cues across frequency bands~\cite{spatialhear1}. Motivated by this, we decompose spectrograms into three frequency resolutions: fine bands (32-bin width), mid bands (128-bin width), and the full spectrum. For each sub-band, we compute Interaural Phase Difference (IPD) and Interaural Level Difference (ILD):
\begin{equation}
\mathrm{IPD}^b(t,f)
=
\angle X^b_1(t,f)\;-\;\angle X^b_2(t,f)
\end{equation}
\begin{equation}
\mathrm{ILD}^b(t,f)
= 
20 \,\log_{10}\!\biggl(
\frac{\bigl|X^b_1(t,f)\bigr|}
     {\bigl|X^b_2(t,f)\bigr| + \epsilon}
\biggr)
\end{equation}
where b denotes the band type (fine, mid, or full), $\epsilon$ is a small constant for stability, $\angle X$ represents the phase, and $\bigl|X\bigr|$ the magnitude of the spectrogram. This process yields the tri-band spatial features ${\bf X}^{\text{fine}}\in \mathbb{R}^{2\times T \times \left \lfloor  \frac{F}{32}\right \rfloor \times 32}$, ${\bf X}^{\text{mid}}\in \mathbb{R}^{2\times T \times \left \lfloor  \frac{F}{128}\right \rfloor \times 128}$, and ${\bf X}^{\text{full}}\in \mathbb{R}^{2\times T \times F\times1}$.

\noindent \textbf{SSA block.} Leveraging the SBF rich in target semantics, we design the SSA block to selectively isolate the spatial characteristics of the target sound source. As depicted in Figure 2(b), the SSA block comprises two key components: a Semantic-Spatial Cross-Attention module and a Semantic-Guided Band-wise Fusion module.

\begin{table*}[!htbp]
\centering
\label{tab_fwsc}
\begin{tabular}{lccccccc}
\toprule
\multirow{2}{*}{\textbf{Model}} & \multirow{2}{*}{\textbf{Modality}} & \multicolumn{2}{c}{0 dB} & \multicolumn{2}{c}{-5 dB} & \multicolumn{2}{c}{-10 dB} \\
\cmidrule(lr){3-4} \cmidrule(lr){5-6} \cmidrule(lr){7-8}
 &  & MAE($^{\circ}$) $\downarrow$ & ACC(\%) $\uparrow$ & MAE($^{\circ}$) $\downarrow$ & ACC(\%) $\uparrow$ & MAE($^{\circ}$) $\downarrow$ & ACC(\%) $\uparrow$ \\
\midrule
SELDNet & Audio Only & 32.19 & 51.27 & 32.82 & 49.91 & 33.49 & 44.67 \\
GCC-MLP & Audio Only & 30.08 & 54.89 & 30.94 & 52.98 & 31.22 & 51.69 \\
\midrule
MLP-AVC & Audio-Visual & 28.92 & 55.28 & 29.13 & 51.63 & 29.97 & 50.74 \\
AVMLP & Audio-Visual & 25.42 & 59.34 & 30.07 & 55.24 & 31.72 & 51.62 \\
DGB & Audio-Visual & 19.09 & 63.87 & 30.38 & 56.63 & 30.98 & 52.87\\
AVST & Audio-Visual & 21.53 & 64.14 & 26.94 & 58.71 & 27.21 & 55.51 \\
AVSELD & Audio-Visual & 18.95 & 67.56 & 22.81 & 60.86 & 24.69 & 57.21 \\
CMAF & Audio-Visual & 18.65 & 67.71 & 22.17 & 61.90 & 24.52 & 57.51 \\
\bf AV-SSAN (Ours) & Audio-Visual & \bf 16.59 & \bf 71.29 & \bf 19.77 & \bf 65.28 & \bf 23.08 & \bf 60.19 \\
\bottomrule
\end{tabular}
\caption{Experimental results under different SNRs on VGG-SSL.}
\label{table2}
\end{table*}

For each sub-band spatial feature $\mathbf{X}^b_{:, i} \in \mathbb{R}^{2 \times T \times d_b}$, where $b$ denotes the band type (fine, mid, or full), $i = 1, \dots, n$ represents the patch index, and $d_b$ is the band width, we apply shared convolutional layers to encode it into $\mathbf{X}^{'b}_{:, i} \in \mathbb{R}^{2 \times T \times 64}$. The encoded feature $\mathbf{X}^{'b}_{:, i }$ is then processed through a cross-attention mechanism, with the SBF serving as the query to extract target-related spatial information:
\begin{equation}
\mathbf{q}^b_i = \mathbf{W}^q_{b,i} \text{SBF}, \quad \mathbf{k}^b_i = \mathbf{W}^k_{b,i} \mathbf{X}^{'b}_{:,i}, \quad \mathbf{v}^b_i = \mathbf{W}^v_{b,i} \mathbf{X}^{'b}_{:,i},
\end{equation}
where $\mathbf{W}^q_{b,i}$, $\mathbf{W}^k_{b,i}$, and $\mathbf{W}^v_{b,i}$ are learnable projection matrices. The attention output is computed as:
\begin{equation}
\mathbf{Z}^b_i = \text{softmax}\left( \frac{\mathbf{q}^b_i (\mathbf{k}^b_i)^\top}{\sqrt{D}} \right) \mathbf{v}^b_i,
\end{equation}
where $D$ is the feature dimension.

Different sound events often exhibit energy concentration in distinct frequency ranges (e.g., whispers in high frequencies, bass in low). Motivated by this observation, we hypothesize that target-specific spatial characteristics may similarly vary across frequency bands, and that their semantic cross-attention weights should reflect this trend. To exploit this frequency-dependent behavior, we introduce a Semantic-Guided Band-wise Fusion module, which adaptively aggregates sub-band features based on their semantic relevance. This allows the model to assign greater weights to the sub-band features that are most informative for the given target. Specifically, we compute a semantic attention map for each patch as:
\begin{equation}
\mathbf{A}^b_i = \mathbf{q}^b_i (\mathbf{k}^b_i)^\top.
\end{equation}
These attention maps are concatenated to form $\mathbf{A}^b = [\mathbf{A}^b_1, \dots, \mathbf{A}^b_n]$. A scalar band importance vector $\boldsymbol{\beta}^b = [\beta^b_1, \dots, \beta^b_n]$ is derived by applying mean pooling and softmax over $\mathbf{A}^b$. The final aggregated feature is computed as:
\begin{equation}
\mathbf{Z}^b = \sum_{i=1}^n \beta^b_i \mathbf{Z}^b_i.
\end{equation}

\noindent\textbf{Spatial Refiner.} To unify the tri-band features $[\mathbf{Z}^{\text{fine}}, \mathbf{Z}^{\text{mid}}, \mathbf{Z}^{\text{full}}] \in \mathbb{R}^{3 \times T \times 64}$, we employ an MLP that predicts temporal importance scores across bands. After softmax normalization, we compute a weighted sum:
\begin{equation}
\mathbf{Z}' = \sum_{b \in \{\text{fine}, \text{mid}, \text{full}\}} \alpha^b(t) \cdot \mathbf{Z}^b(t)
\end{equation}
The fused feature $\mathbf{Z}' \in \mathbb{R}^{T \times 64}$ is passed through an MLP to predict a TF mask:
\begin{equation}
\mathbf{F}_{\text{Mask}} = \text{MLP}(\mathbf{Z}')
\end{equation}
This mask is applied to the original spectrogram:
\begin{equation}
\mathbf{X}m = \mathbf{X} \otimes \mathbf{F}_{\text{Mask}}
\end{equation}
where $\otimes$ denotes element-wise multiplication.

\subsection{DoA Prediction}
Using the masked spectrogram $\mathbf{X}_m$ as input, an MLP followed by a softmax layer predicts the DoA posterior:
\begin{equation}
\hat{p}(\theta) = \text{Softmax}(\text{MLP}(\mathbf{X}_m))
\end{equation}

\subsection{Training Loss}
To optimize reconstruction loss between the masked spectrogram ${\bf X}_{m}$ and the ground truth ${\bf X}_{gt}$, we adopt a Mean Squared Error (MSE) loss:
\begin{equation}
    \mathcal{L}_{recon} = \left \| {\bf X}_{m} - {\bf X}_{gt} \right \|^{2}_{2} 
\end{equation}

An MSE loss is used for posterior probability-based DoA estimation, optimizing the predicted DoA distribution:
\begin{equation}
    \mathcal{L}_{DoA} = \sum_{\theta=1}^{180}\left \| \hat{p}(\theta)-p(\theta)\right \|^{2}_{2} 
\end{equation}

The final joint loss function is formulated as:
\begin{equation}
    \arg\min \mathcal{L} = \mathcal{L}_{recon} + \mathcal{L}_{DoA}
\end{equation}

\begin{figure*}[!htb]
    \centering
    \includegraphics[width=0.75\linewidth]{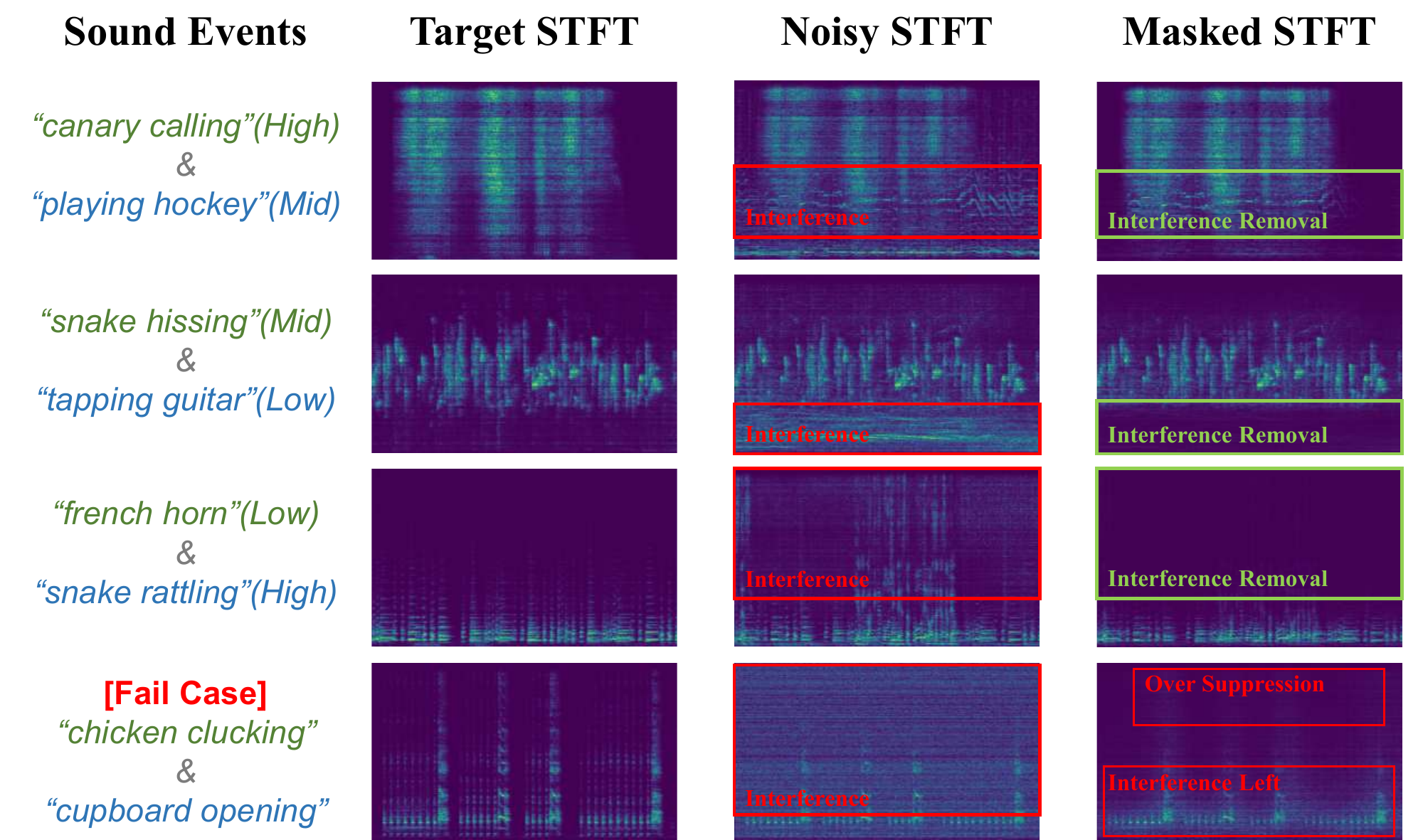}
    \caption{Visualizations of AV-SSAN's outputs on VGG-SSL. Each row shows a pair of overlapping sound events categorized by their dominant frequency regions: high (H), mid (M), or low (L).}
    \label{mask}
\end{figure*}
\section{Experiments \& Discussions}
\subsection{Dataset Construction}
Table~\ref{table1} provides a review of existing AV-SSL datasets. AV16.3~\cite{AV16.3}, CAV3D~\cite{cav3d}, SSLR~\cite{gccmlp}, and AVRI~\cite{CMAF} offer 3D location annotations with real-recorded spatial audio. However, they primarily contain human speech, restricting their utility in more general localization tasks involving diverse sound events. While STARSS23~\cite{starss23} includes 13 sound events with spatial annotations, its prevalence of overlapping, speech-dominated sources renders it less suitable for evaluating disentangled audio-visual spatial localization.

Our proposed VGGSound-SSL dataset, derived from VGGSound~\cite{vggsound}, comprises two-channel spatial audio and semantically aligned visual prompts for 296 distinct sound events. Its construction pipeline involves two primary stages—spatial audio synthesis and prompt image generation (detailed pipeline see \textit{Appendix B}).

\noindent\textbf{Spatial Audio Synthesis.} We extract 10-second single-channel audio clips from VGGSound videos and resample them to 16 kHz. To simulate spatial audio, each audio segment is convolved with a randomly selected room impulse response (RIR) using GPU-RIR~\cite{gpurir}. A total of 10,000 RIRs are synthesized by varying critical parameters, including room dimensions, sound source positions, and reverberation times (T60). The distribution of these parameters is provided in \textit{Appendix B}.

\noindent\textbf{Prompt Image Generation.}
To provide semantically consistent visual cues, we extract frames from each video and compute their CLIP embeddings. Concurrently, text embeddings for each sound class are obtained using CLIP. The frame exhibiting the highest image-text similarity is then selected as the prompt image, thus ensuring strong semantic alignment between the sound source and its visual reference. All selected prompt images were manually verified to ensure semantic consistency and reduce selection bias.


All selected prompt images are organized into class-specific pools. During training and inference, a prompt image of the target sound is randomly sampled from its corresponding pool, excluding the image from the current instance.

\begin{table}[!tb]
    \centering
    \begin{tabular}{ccc}
    \toprule
    \textbf{Model} & MAE($^{\circ}$) & ACC(\%)\\
    \midrule
    MLP-AVC & 36.85 & 41.23\\
    AVMLP & 35.19 & 40.12 \\
    DGB & 36.27 & 42.76 \\
    AVST & 33.28 & 45.34\\
    AVSELD & 28.54 & 51.64 \\
    CMAF & 30.03 & 50.87 \\
    \bf AV-SSAN (Ours)& \bf 27.46 & \bf 52.31\\
    \bottomrule
    \end{tabular}
    \caption{Experimental results on STARSS23.}
    \label{real_data}
\end{table}

\subsection{Baselines}
We compare our proposed method with
audio-only baselines, including 1) SELDNet~\cite{seldnet} and 2) GCC-MLP~\cite{gccmlp}, which estimate DoA using only the mixed spatial audio signals. For audio-visual methods, we evaluate 3) MLP-AVC~\cite{qianmulti}, 4) AVMLP~\cite{avmlp}, 5) DGB~\cite{dgb}, 6) AVST~\cite{AVST}, 7) AVSELD~\cite{avseld}, and 8) CMAF~\cite{CMAF}.
The implementation details of each method are given in \textit{Appendix C}.

\subsection{Training and Evaluation}
We train all models on the proposed VGGSound-SSL dataset. Each training sample is formed by randomly selecting two audio segments from distinct sound classes. One is selected as the target source, and the other as the interfering source. The two segments are mixed at an SNR of 0 dB. For visual prompting, we randomly selected an image corresponding to a different instance within the same category.

\begin{figure*}[!htb]
    \centering
    \includegraphics[width=0.8\linewidth]{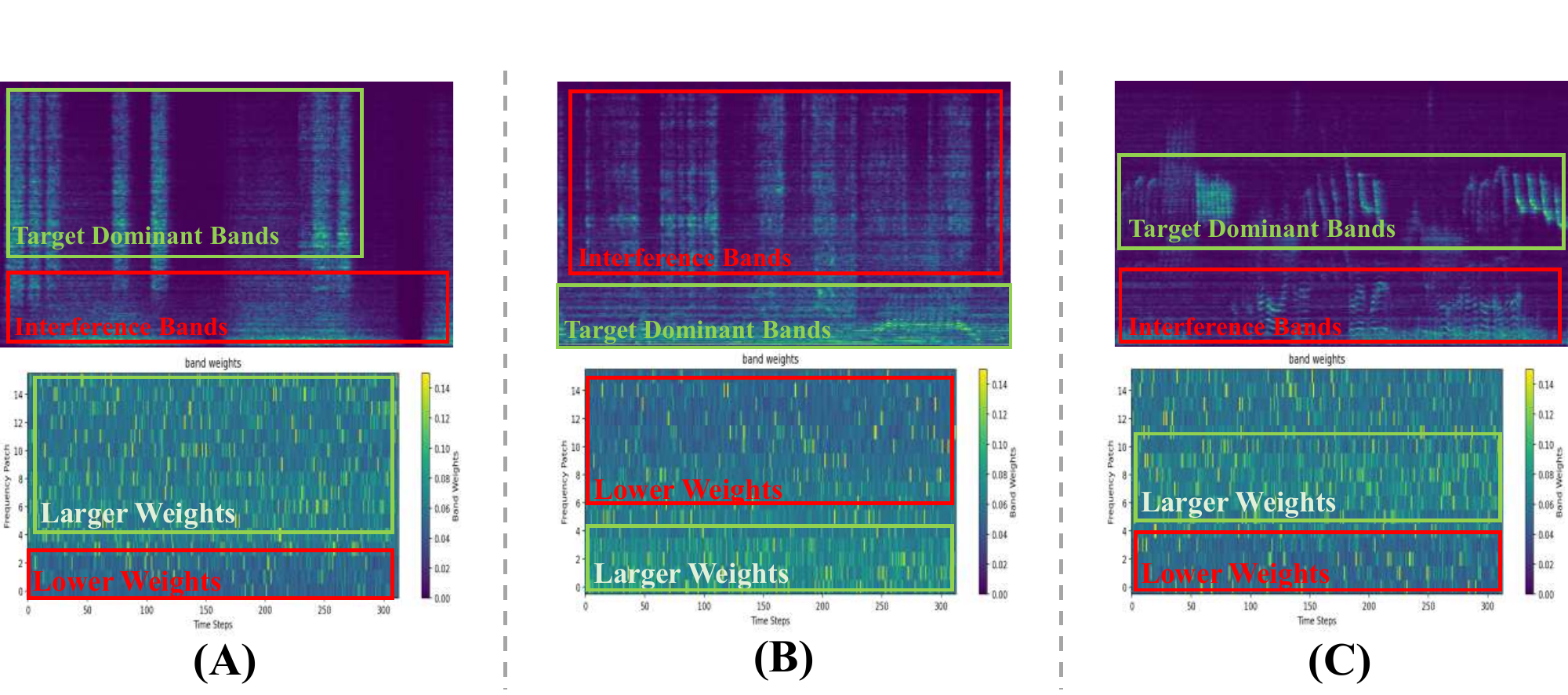}
    \caption{Visualizations of the ablation study on band attention. Our model assigns higher weights to frequency bands dominated by the target and suppresses interference-dominated bands.}
    \label{band weight}
\end{figure*}

To assess generalization, we further train models on the STARSS23 dataset, which contains real-world spatial audio recordings. For each 10-second segment, the most frequent sound event is selected as the target, with remaining events treated as interference. Since STARSS23 lacks visual counterparts, we use category-matched images from VGGSound-SSL as semantic prompts.

We evaluate performance using Mean Absolute Error (MAE) and Accuracy (ACC). MAE quantifies the average angular error, while ACC reports the percentage of predictions within a $5^\circ$ tolerance. Lower MAE and higher ACC indicate stronger localization capability.

All the experiments are conducted on two RTX-4090 GPUs. Our model was trained for 140k steps with a batch size of 32, using the AdamW optimizer with a learning rate of 5e-3. For STFT computations, a frame size of 64 ms and a hop size of 32 ms were employed.




\subsection{Results and Analysis}
Table~\ref{table2} presents performance comparisons on VGG-SSL across varying SNR levels. At SNR = 0 dB, our proposed AV-SSAN achieves the lowest MAE of $16.59^{\circ}$ and the highest ACC of 71.29\%, outperforming all competing methods.

As expected, audio-only models such as SELDNet and GCC-MLP yield ACC near 50\%. This aligns with the binary choice setting of the task: without visual guidance, the model must choose between two plausible sources. The result confirms the necessity of visual guidance to resolve spatial ambiguity. 

Compared to prior audio-visual methods, AV-SSAN shows superior accuracy. This validates the effectiveness of our Multi-band Semantic-Spatial Alignment framework, which explicitly aligns visual semantics with spatial auditory patterns across frequency bands. 

To evaluate robustness in noisy environments, we conduct a generalization experiment. All models are trained at 0 dB and tested under SNRs ranging from -5 dB to –10 dB. As shown in Table~\ref{table2}, AV-SSAN retains strong performance and outperforms all baselines. This indicates that our method captures discriminative spatial cues resilient to interference, demonstrating strong generalization across noise conditions. 

We further assess transferability on the STARSS23 dataset, which features real-world acoustic environments. As shown in Table~\ref{real_data}, all models experience performance drops due to the increased acoustic complexity of real-world scenes, where multiple overlapping sound events create more challenging interference patterns. AV-SSAN remains the best performer, demonstrating stronger transferability to in-the-wild conditions.

Figure~\ref{mask} provides qualitative examples of VGG-SSL. Our method effectively suppresses interference while preserving the target, showing the benefits of multi-band alignment. We also include a failure case where the target energy is over-suppressed and interference remains. We attribute this to an energy imbalance between the target and interference, which hampers the selective localization.

Computational complexity analysis, prompt selection study, sensitivity analysis, and additional visualizations are provided in the \textit{Appendix D\&F}.
\subsection{Ablation Study}
We conduct ablation studies on the VGG-SSL dataset to isolate the contribution of each component in our proposed AV-SSAN framework. Results are summarized in Table~\ref{ablation}.

\begin{table}
    \centering
    \begin{tabular}{ccccccc}
    \toprule
    V & A & MB & BA & Refiner & MAE ($ ^\circ$) & ACC (\%) \\
    \midrule
    $\surd$ & $\times$ & $\times$ & $\times$ & $\times$ & 28.54 & 51.38 \\
    $\surd$ & $\surd$ & $\times$ & $\times$ & $\times$ & 22.21 & 61.99\\
    $\surd$ & $\surd$ & $\surd$ & $\times$ & $\times$ & 19.19 & 68.04\\
    $\surd$ & $\surd$ & $\surd$ & $\surd$ & $\times$ & 17.98 & 69.48 \\
    $\surd$ & $\surd$ & $\surd$ & $\surd$ & $\surd$ & 16.59 & 71.29 \\
    \bottomrule
\end{tabular}
\caption{Ablation study on VGG-SSL. V: visual prompt; A: semantic audio; MB: multi-band modeling; BA: band attention; Refiner: spatial refiner.}
\label{ablation}
\end{table}
\noindent\textbf{Explicit Semantic-Spatial Alignment.} We first evaluate the importance of semantic and spatial alignment. Simply applying cross-attention between cross-instance visual prompts and mixed spatial audio yields near-random performance of 51.38\%. These results suggest a strong modality mismatch between the two modalities. Incorporating Semantic Audio representations reduces ambiguity, improving ACC by 10.61\% and lowering MAE by $6.33^{\circ}$. It confirms that bridging visual semantics and spatial acoustics at a semantic level is critical for disambiguating the target.

\noindent\textbf{Multi-band Semantic-Spatial Alignment.} Adding multi-band modeling further improves performance, increasing ACC by 6.05\%. This validates our design to align spatial features across multiple frequency bands. By decomposing the audio into coarse-to-fine frequency resolutions, the model better captures target-specific spatial patterns that may be frequency-dependent.

\noindent\textbf{Band Attention.} Introducing Band Attention yields additional gains, with ACC reaching 69.48\% and the MAE dropping to $17.98^{\circ}$. This confirms that not all frequency bands contribute equally. Our attention mechanism allows the model to prioritize sub-bands that are more discriminative for a given target class, thus enhancing localization performance.

Figure 4 provides a visualization of the learned band attentions. Patches corresponding to the target exhibit stronger responses, while interference regions are suppressed. This validates our hypothesis: semantic-spatial cross attention should reflect frequency-specific discriminability, and band attention enables this behavior to emerge dynamically.

\noindent\textbf{Spatial Refiner.} Finally, the addition of the Spatial Refiner leads to further improvements of 1.8\% in ACC and $1.39^{\circ}$ in MAE. This module integrates refined spatial features from attended sub-bands, enabling the model to make more globally consistent predictions based on the band-wise aligned information.


\section{Conclusion}
We introduce CI-AVL, a novel and challenging task that localizes a target sound source using semantically related but spatially unpaired visual prompts. To solve this task, we propose AV-SSAN, a selective localization framework that performs multi-band semantic-spatial alignment between visual semantics and spatial audio cues. To support research in this direction, we construct VGG-SSL, a large-scale dataset with 13,981 spatial audio clips and class-consistent prompt images. Extensive experiments show that AV-SSAN achieves state-of-the-art performance.


\bibliography{aaai2026}

\begin{thebibliography}{39}
\providecommand{\natexlab}[1]{#1}

\bibitem[{Adavanne et~al.(2018)Adavanne, Politis, Nikunen, and Virtanen}]{seldnet}
Adavanne, S.; Politis, A.; Nikunen, J.; and Virtanen, T. 2018.
\newblock Sound event localization and detection of overlapping sources using convolutional recurrent neural networks.
\newblock \emph{IEEE Journal of Selected Topics in Signal Processing}, 13(1): 34--48.

\bibitem[{Berghi et~al.(2024)Berghi, Wu, Zhao, Wang, and Jackson}]{avseld}
Berghi, D.; Wu, P.; Zhao, J.; Wang, W.; and Jackson, P.~J. 2024.
\newblock Fusion of audio and visual embeddings for sound event localization and detection.
\newblock In \emph{IEEE International Conference on Acoustics, Speech and Signal Processing}, 8816--8820. IEEE.

\bibitem[{Brughera, Dunai, and Hartmann(2013)}]{spatialhear2}
Brughera, A.; Dunai, L.; and Hartmann, W.~M. 2013.
\newblock Human interaural time difference thresholds for sine tones: The high-frequency limit.
\newblock \emph{The Journal of the Acoustical Society of America}, 133(5): 2839--2855.

\bibitem[{Chakrabarty and Habets(2017)}]{cnndoa}
Chakrabarty, S.; and Habets, E.~A. 2017.
\newblock Broadband DOA estimation using convolutional neural networks trained with noise signals.
\newblock In \emph{IEEE Workshop on Applications of Signal Processing to Audio and Acoustics}, 136--140. IEEE.

\bibitem[{Chen et~al.(2020)Chen, Xie, Vedaldi, and Zisserman}]{vggsound}
Chen, H.; Xie, W.; Vedaldi, A.; and Zisserman, A. 2020.
\newblock Vggsound: A large-scale audio-visual dataset.
\newblock In \emph{IEEE International Conference on Acoustics, Speech and Signal Processing}, 721--725. IEEE.

\bibitem[{Chen et~al.(2024)Chen, Qian, Pan, Chen, and Li}]{locselect}
Chen, Y.; Qian, X.; Pan, Z.; Chen, K.; and Li, H. 2024.
\newblock LocSelect: Target Speaker Localization with an Auditory Selective Hearing Mechanism.
\newblock In \emph{IEEE International Conference on Acoustics, Speech and Signal Processing}, 8696--8700. IEEE.

\bibitem[{Diaz-Guerra, Miguel, and Beltran(2021)}]{gpurir}
Diaz-Guerra, D.; Miguel, A.; and Beltran, J.~R. 2021.
\newblock gpuRIR: A python library for room impulse response simulation with GPU acceleration.
\newblock \emph{Multimedia Tools and Applications}, 80(4): 5653--5671.

\bibitem[{DiBiase, Silverman, and Brandstein(2001)}]{SRPPHAT}
DiBiase, J.~H.; Silverman, H.~F.; and Brandstein, M.~S. 2001.
\newblock Robust localization in reverberant rooms.
\newblock In \emph{Microphone arrays: signal processing techniques and applications}, 157--180. Springer.

\bibitem[{Dosovitskiy et~al.(2020)Dosovitskiy, Beyer, Kolesnikov, Weissenborn, Zhai, Unterthiner, Dehghani, Minderer, Heigold, Gelly et~al.}]{vit}
Dosovitskiy, A.; Beyer, L.; Kolesnikov, A.; Weissenborn, D.; Zhai, X.; Unterthiner, T.; Dehghani, M.; Minderer, M.; Heigold, G.; Gelly, S.; et~al. 2020.
\newblock An image is worth 16x16 words: Transformers for image recognition at scale.
\newblock \emph{arXiv preprint arXiv:2010.11929}.

\bibitem[{Gulati et~al.(2020)Gulati, Qin, Chiu, Parmar, Zhang, Yu, Han, Wang, Zhang, Wu et~al.}]{conformer}
Gulati, A.; Qin, J.; Chiu, C.-C.; Parmar, N.; Zhang, Y.; Yu, J.; Han, W.; Wang, S.; Zhang, Z.; Wu, Y.; et~al. 2020.
\newblock Conformer: Convolution-augmented transformer for speech recognition.
\newblock \emph{arXiv preprint arXiv:2005.08100}.

\bibitem[{He, Motlicek, and Odobez(2018)}]{gccmlp}
He, W.; Motlicek, P.; and Odobez, J.-M. 2018.
\newblock Deep neural networks for multiple speaker detection and localization.
\newblock In \emph{IEEE International Conference on Robotics and Automation}, 74--79. IEEE.

\bibitem[{Hershey et~al.(2017)Hershey, Chaudhuri, Ellis, Gemmeke, Jansen, Moore, Plakal, Platt, Saurous, Seybold et~al.}]{vggish}
Hershey, S.; Chaudhuri, S.; Ellis, D.~P.; Gemmeke, J.~F.; Jansen, A.; Moore, R.~C.; Plakal, M.; Platt, D.; Saurous, R.~A.; Seybold, B.; et~al. 2017.
\newblock CNN architectures for large-scale audio classification.
\newblock In \emph{IEEE International Conference on Acoustics, Speech and Signal Processing}, 131--135. IEEE.

\bibitem[{Jiang, Han, and Mesgarani(2025)}]{bimamba}
Jiang, X.; Han, C.; and Mesgarani, N. 2025.
\newblock Dual-path mamba: Short and long-term bidirectional selective structured state space models for speech separation.
\newblock In \emph{IEEE International Conference on Acoustics, Speech and Signal Processing}, 1--5. IEEE.

\bibitem[{Jiang et~al.(2023)Jiang, Chen, Du, Wang, and Lee}]{lipnet}
Jiang, Y.; Chen, H.; Du, J.; Wang, Q.; and Lee, C.-H. 2023.
\newblock Incorporating lip features into audio-visual multi-speaker doa estimation by gated fusion.
\newblock In \emph{IEEE International Conference on Acoustics, Speech and Signal Processing}, 1--5. IEEE.

\bibitem[{Knapp and Carter(1976)}]{GCC}
Knapp, C.; and Carter, G. 1976.
\newblock The generalized correlation method for estimation of time delay.
\newblock \emph{IEEE Transactions on acoustics, speech, and signal processing}, 24(4): 320--327.

\bibitem[{Lathoud, Odobez, and Gatica-Perez(2004)}]{AV16.3}
Lathoud, G.; Odobez, J.-M.; and Gatica-Perez, D. 2004.
\newblock AV16. 3: An audio-visual corpus for speaker localization and tracking.
\newblock In \emph{International Workshop on Machine Learning for Multimodal Interaction}, 182--195. Springer.

\bibitem[{Li, Zhang, and Li(2018)}]{cnnlstm_doa}
Li, Q.; Zhang, X.; and Li, H. 2018.
\newblock Online direction of arrival estimation based on deep learning.
\newblock In \emph{IEEE International Conference on Acoustics, Speech and Signal Processing}, 2616--2620. IEEE.

\bibitem[{Pertil{\"a} and Cakir(2017)}]{srp2}
Pertil{\"a}, P.; and Cakir, E. 2017.
\newblock Robust direction estimation with convolutional neural networks based steered response power.
\newblock In \emph{IEEE International Conference on Acoustics, Speech and Signal Processing}, 6125--6129. IEEE.

\bibitem[{Qian et~al.(2019)Qian, Brutti, Lanz, Omologo, and Cavallaro}]{cav3d}
Qian, X.; Brutti, A.; Lanz, O.; Omologo, M.; and Cavallaro, A. 2019.
\newblock Multi-speaker tracking from an audio--visual sensing device.
\newblock \emph{IEEE Transactions on Multimedia}, 21(10): 2576--2588.

\bibitem[{Qian et~al.(2021{\natexlab{a}})Qian, Brutti, Lanz, Omologo, and Cavallaro}]{avmlp}
Qian, X.; Brutti, A.; Lanz, O.; Omologo, M.; and Cavallaro, A. 2021{\natexlab{a}}.
\newblock Audio-visual tracking of concurrent speakers.
\newblock \emph{IEEE Transactions on Multimedia}, 24: 942--954.

\bibitem[{Qian et~al.(2021{\natexlab{b}})Qian, Madhavi, Pan, Wang, and Li}]{qianmulti}
Qian, X.; Madhavi, M.; Pan, Z.; Wang, J.; and Li, H. 2021{\natexlab{b}}.
\newblock Multi-target DoA estimation with an audio-visual fusion mechanism.
\newblock In \emph{IEEE International Conference on Acoustics, Speech and Signal Processing}, 4280--4284. IEEE.

\bibitem[{Qian et~al.(2022{\natexlab{a}})Qian, Wang, Wang, Guan, and Li}]{CMAF}
Qian, X.; Wang, Z.; Wang, J.; Guan, G.; and Li, H. 2022{\natexlab{a}}.
\newblock Audio-visual cross-attention network for robotic speaker tracking.
\newblock \emph{IEEE/ACM Transactions on Audio, Speech, and Language Processing}, 31: 550--562.

\bibitem[{Qian et~al.(2022{\natexlab{b}})Qian, Zhang, Guan, and Xue}]{dgb}
Qian, X.; Zhang, Q.; Guan, G.; and Xue, W. 2022{\natexlab{b}}.
\newblock Deep audio-visual beamforming for speaker localization.
\newblock \emph{IEEE Signal Processing Letters}, 29: 1132--1136.

\bibitem[{Radford et~al.(2021)Radford, Kim, Hallacy, Ramesh, Goh, Agarwal, Sastry, Askell, Mishkin, Clark et~al.}]{CLIP}
Radford, A.; Kim, J.~W.; Hallacy, C.; Ramesh, A.; Goh, G.; Agarwal, S.; Sastry, G.; Askell, A.; Mishkin, P.; Clark, J.; et~al. 2021.
\newblock Learning transferable visual models from natural language supervision.
\newblock In \emph{International Conference on Machine Learning}, 8748--8763. PmLR.

\bibitem[{Schmidt(1986)}]{MUSIC}
Schmidt, R. 1986.
\newblock Multiple emitter location and signal parameter estimation.
\newblock \emph{IEEE Transactions on Antennas and Propagation}, 34(3): 276--280.

\bibitem[{Shimada et~al.(2023)Shimada, Politis, Sudarsanam, Krause, Uchida, Adavanne, Hakala, Koyama, Takahashi, Takahashi et~al.}]{starss23}
Shimada, K.; Politis, A.; Sudarsanam, P.; Krause, D.~A.; Uchida, K.; Adavanne, S.; Hakala, A.; Koyama, Y.; Takahashi, N.; Takahashi, S.; et~al. 2023.
\newblock STARSS23: An audio-visual dataset of spatial recordings of real scenes with spatiotemporal annotations of sound events.
\newblock \emph{Advances in Neural Information Processing Systems}, 36: 72931--72957.

\bibitem[{Shimada et~al.(2024)Shimada, Uchida, Koyama, Shibuya, Takahashi, Mitsufuji, and Kawahara}]{zerolocate}
Shimada, K.; Uchida, K.; Koyama, Y.; Shibuya, T.; Takahashi, S.; Mitsufuji, Y.; and Kawahara, T. 2024.
\newblock Zero-and few-shot sound event localization and detection.
\newblock In \emph{IEEE International Conference on Acoustics, Speech and Signal Processing}, 636--640. IEEE.

\bibitem[{Shmuel et~al.(2023)Shmuel, Merkofer, Revach, Van~Sloun, and Shlezinger}]{music1}
Shmuel, D.~H.; Merkofer, J.~P.; Revach, G.; Van~Sloun, R.~J.; and Shlezinger, N. 2023.
\newblock Deep root MUSIC algorithm for data-driven DoA estimation.
\newblock In \emph{IEEE International Conference on Acoustics, Speech and Signal Processing}, 1--5. IEEE.

\bibitem[{Slizovskaia et~al.(2022)Slizovskaia, Wichern, Wang, and Le~Roux}]{classlocate}
Slizovskaia, O.; Wichern, G.; Wang, Z.-Q.; and Le~Roux, J. 2022.
\newblock Locate this, not that: Class-conditioned sound event doa estimation.
\newblock In \emph{IEEE International Conference on Acoustics, Speech and Signal Processing}, 711--715. IEEE.

\bibitem[{Strutt(1907)}]{spatialhear1}
Strutt, J.~W. 1907.
\newblock On our perception of sound direction.
\newblock \emph{Philosophical Magazine}, 13(74): 214--32.

\bibitem[{Taevs et~al.(2010)Taevs, Dahmani, Zatorre, and Bohbot}]{semantic1}
Taevs, M.; Dahmani, L.; Zatorre, R.~J.; and Bohbot, V.~D. 2010.
\newblock Semantic elaboration in auditory and visual spatial memory.
\newblock \emph{Frontiers in Psychology}, 1: 228.

\bibitem[{van~der Heijden et~al.(2019)van~der Heijden, Rauschecker, de~Gelder, and Formisano}]{semantic2}
van~der Heijden, K.; Rauschecker, J.~P.; de~Gelder, B.; and Formisano, E. 2019.
\newblock Cortical mechanisms of spatial hearing.
\newblock \emph{Nature Reviews Neuroscience}, 20(10): 609--623.

\bibitem[{Vaswani et~al.(2017)Vaswani, Shazeer, Parmar, Uszkoreit, Jones, Gomez, Kaiser, and Polosukhin}]{transformer}
Vaswani, A.; Shazeer, N.; Parmar, N.; Uszkoreit, J.; Jones, L.; Gomez, A.~N.; Kaiser, {\L}.; and Polosukhin, I. 2017.
\newblock Attention is all you need.
\newblock \emph{Advances in Neural Information Processing Systems}, 30.

\bibitem[{Wang, Yang, and Li(2024)}]{spatialnet}
Wang, Y.; Yang, B.; and Li, X. 2024.
\newblock IPDnet: A universal direct-path IPD estimation network for sound source localization.
\newblock \emph{IEEE/ACM Transactions on Audio, Speech, and Language Processing}.

\bibitem[{Wu, Hu, and Wang(2023)}]{MultiDoA}
Wu, Y.; Hu, R.; and Wang, X. 2023.
\newblock Multi-speaker Direction of Arrival Estimation Using Audio and Visual Modalities with Convolutional Neural Network.
\newblock In \emph{IEEE International Conference on Multimedia and Expo}, 636--641. IEEE.

\bibitem[{Wu et~al.(2023)Wu, Hu, Wang, and Ke}]{TAVF}
Wu, Y.; Hu, R.; Wang, X.; and Ke, S. 2023.
\newblock Multi-speaker DoA estimation using audio and visual modality.
\newblock \emph{Neural Processing Letters}, 55(7): 8887--8901.

\bibitem[{Xiao and Das(2024)}]{tfmamba}
Xiao, Y.; and Das, R.~K. 2024.
\newblock Tf-mamba: A time-frequency network for sound source localization.
\newblock \emph{arXiv preprint arXiv:2409.05034}.

\bibitem[{Zhao et~al.(2024)Zhao, Qian, Xu, Liu, Cao, Berghi, and Wang}]{textlocate}
Zhao, J.; Qian, X.; Xu, Y.; Liu, H.; Cao, Y.; Berghi, D.; and Wang, W. 2024.
\newblock Text-queried target sound event localization.
\newblock In \emph{European Signal Processing Conference}, 261--265. IEEE.

\bibitem[{Zhao et~al.(2023)Zhao, Xu, Qian, and Wang}]{AVST}
Zhao, J.; Xu, Y.; Qian, X.; and Wang, W. 2023.
\newblock Audio visual speaker localization from egocentric views.
\newblock \emph{arXiv preprint arXiv:2309.16308}.

\end{thebibliography}

\end{document}